\documentclass[11pt]{article}

\date{}

\usepackage[utf8]{inputenc} 
\usepackage[T1]{fontenc}    
\usepackage{hyperref}       
\usepackage{url}            
\usepackage{booktabs}       
\usepackage{amsfonts}       
\usepackage{nicefrac}       
\usepackage{microtype}      
\usepackage{xspace}

\usepackage{comment}
\usepackage{amsmath}
\usepackage{amsthm}
\usepackage{graphicx}
\usepackage{rotating}

\usepackage{amssymb}
\usepackage{autobreak}
\usepackage{mathtools}
\usepackage{xcolor}
\usepackage{bbm}
\usepackage{algorithm}
\usepackage{algorithmic}

\makeatletter

\title{EikoNet: Solving the Eikonal equation with Deep Neural Networks}

\author{Jonathan D. Smith \footnote{Seismological Laboratory, California Institute of Technology, Pasadena, CA, USA} , Kamyar Azizzadenesheli \footnote{Department of Computing and Mathematical Sciences, California Institute of Technology, Pasadena, CA, USA} , Zachary E. Ross \footnotemark[1]}

\begin{document}

\maketitle

\begin{abstract}
The recent deep learning revolution has created an enormous opportunity for accelerating compute capabilities in the context of physics-based simulations. Here, we propose EikoNet, a deep learning approach to solving the Eikonal equation, which characterizes the first-arrival-time field in heterogeneous 3D velocity structures. Our grid-free approach allows for rapid determination of the travel time between any two points within a continuous 3D domain. These travel time solutions are allowed to violate the differential equation---which casts the problem as one of optimization---with the goal of finding network parameters that minimize the degree to which the equation is violated. In doing so, the method exploits the differentiability of neural networks to calculate the spatial gradients analytically, meaning the network can be trained on its own without ever needing solutions from a finite difference algorithm. EikoNet is rigorously tested on several velocity models and sampling methods to demonstrate robustness and versatility. Training and inference are highly parallelized, making the approach well-suited for GPUs. EikoNet has low memory overhead, and further avoids the need for travel-time lookup tables. The developed approach has important applications to earthquake hypocenter inversion, ray multi-pathing, and tomographic modeling, as well as to other fields beyond seismology where ray tracing is essential. 
\end{abstract}

\section{Introduction}
Three-dimensional ray tracing is a fundamental component of modern seismology, having direct applications to earthquake hypocenter inversions \cite{Hauksson2012}, seismic tomography \cite{Zhang2009}, and earthquake source properties \cite{Das1981}. These derived products further form the basis for many downstream seismological applications. The Eikonal equation is a well-known nonlinear partial differential equation that characterizes the first-arrival-time field for a given source location in a 3D medium \cite{Noack2017}. The Eikonal formulation can be solved with several finite difference algorithms \cite{Vidale1988,Podvin1991,Rawlinson2004}, with varying computational demands and stabilities to the solutions.

Recent advances in deep learning have shown to be extremely promising in the context of physics-based simulations \cite{Guo2016,Zhu2018}. This technology has started to be applied to geophysics as well, for example, to predict the acoustic wave response of a medium given a velocity model as input \cite{Moseley2018}, forecast the next time step of a wavefield conditional on its history \cite{Moseley2018}, and accelerate viscoelastic simulations \cite{DeVries2017}. However, these techniques rely on inputs from pre-computed physics based models, which could themselves contain modeling-based artifacts and input bias. Instead, we wish to learn the underlying physics by incorporating the formulations into the neural network architecture and loss function. These physics-based procedures \cite{Rudy2017,Weinan2018,Raissi2019,Bar2019,Li2020} take advantage of the backpropagation procedure to compute the gradient of the neural network output relative to the input terms. Such physics-informed neural networks~\cite{Rudy2017,Raissi2019} (PINNS) are mesh independent, giving a continuous function output related to the inputs. These techniques have been used to learn the parametrization for formulations of the Burgers, Schrodinger, and Navier-Stokes equations, with comparisons made to the numerical derivatives \cite{Rudy2017}.\\

In this paper, we propose EikoNet, an approach to solving the factored Eikonal equation. EikoNet can be trained to learn the travel-time between any two points in a truly continuous 3D medium, avoiding the use of grids. We leverage the differentiability of neural networks to analytically compute the spatial gradients of the travel-time field, and train the network to minimize the difference between the true and predicted velocity model for the factored Eikonal formulation. EikoNet is massively parallelized and therefore well-suited for GPUs, has low memory overhead, and avoids the need for travel-time lookup tables. Additionally, EikoNet has several novel advantages that are not currently offered with conventional finite differencing schemes.

\section{Eikonal Formulation}
The Eikonal equation is a nonlinear first-order partial differential equation representing a high-frequency approximation to the propagation of waves in heterogeneous media \cite{Noack2017}. The equation takes the general form;
\begin{equation}
    \| \nabla{T_{s \rightarrow r}} \| ^2 = \frac{1}{V\left(\vec{x}_r\right)^2} = S\left(\vec{x}_r\right)^2 \label{eq:1}
\end{equation}
where $\|\cdot\|^2$ is the Euclidean norm, $T_{s \rightarrow r}$ is the travel-time through the medium from a source location $s$ to a receiver  location $r$, $V_r$ is the velocity of the medium at the receiver  location, $S_r$ is the slowness of the medium at the receiver location, and $\mathop{\nabla}_{r}$ the gradient at the receiver location.

The value of travel-time is computed by minimizing the misfit of a travel-time field that satisfies the user imposed velocity model, with the additional boundary condition that the travel-time at the source location equals zero, $T_{s \rightarrow s} = 0$. Solutions to equation \ref{eq:1} have a strong singularity at the source location \cite{Treister2016}, leading to numerical errors close to the source. To mitigate such singularity effects, a factored formulation is often used, with solutions representing the travel-time deviation from a homogeneous medium with $V = 1$ \cite{Treister2016}. The factored travel-time form can then be represented by:
\begin{equation}
    T_{s \rightarrow r} = T_0 \cdot \tau_{s \rightarrow r} \label{eq:2}
\end{equation}
where $T_0={\| \vec{x_r} - \vec{x_s} \|}$, representing the distance function from the source location, and $\tau$ the deviation of the travel-time field from a model travel-time with homogeneous unity velocity. Substituting the formulation of equation \ref{eq:2} into equation \ref{eq:1} and expanding using the chain rule, then the velocity can be represented by; \\
\begin{equation}
V\left(\vec{x_r}\right) =  \left[ T_0^2 \| \mathop{\nabla}_{r}\tau_{s\rightarrow r} \| ^2 + 2\tau_{s \rightarrow r}\left(\vec{x_p} - \vec{x_s}\right)\cdot\mathop{\nabla}_{r}\tau_{s \rightarrow r} + \tau_{s \rightarrow r}^2 \right]^{-\frac{1}{2}}\!\!\!\!\!. \label{eq:3}
\end{equation}

The partial differential terms in equation \ref{eq:3} are typically solved using a finite-difference approach and will be discussed in Section \ref{sec:FMM}

\section{Methods}
\subsection{Network architecture and training}
Our approach to solving the Eikonal equation trains a deep neural network, $f_\theta$, to predict the travel-time field, $\tau$, between an input pair of source-receiver coordinates, $\vec{x}=[\vec{x_s}, \vec{x_r}]$. The deviation of the travel-time field is then represented by,
\begin{equation}
    \tau = f_\theta(\vec{x}), \label{eq:4}
\end{equation}
with the corresponding travel-time between source and receiver location represented by equation \ref{eq:2}.

If $\tau$ was known, a neural network could be trained for a catalog source-receiver pairs. However, $\tau$ itself is unknown, being the solution to the equation that we want to solve. Instead, only the velocity model and a differential equation specifying how $\tau$ relates to $V$ are known, but this can be used to train the neural network. Thus, we cast the problem as one where we aim to accurately predict the local velocity, assuming that the output of $f_\theta$ is indeed $\tau$; and use this value to compute $V$ from equation \ref{eq:3}, defining this predicted velocity as $\hat{V}$.  Here, we exploit the differentiability of deep neural networks to analytically determine the spatial gradient of equation \ref{eq:4} with respect to $\vec{x_r}$. This is possible because the layers and activations of the neural network are chosen precisely to be analytically differentiable. In this study, we use Pytorch to calculate the gradients and perform all network training.

Solving the factored Eikonal equation is therefore reduced to training a neural network with supervision on the velocity model, by iteratively updating the parameters $\theta$ to minimize some loss function. Once trained, the Eikonal equation is no longer needed, as $f_\theta$ outputs $\tau$ directly (Figure \ref{fig:FlowDiagram}a). In the process, the details of the velocity model will be encoded in the network, requiring the network to be re-trained if the velocity model is to be changed. As such, the algorithm is fully capable of solving the Eikonal equation from scratch, without needing to run a separate finite-difference simulation to generate training data.

The model architecture is a feed-forward network consisting of a series of residual blocks with fully-connected layers \cite{Huang2017} followed by nonlinear units (Figure \ref{fig:FlowDiagram}a). We use Exponential Linear Unit (ELU) as the activation \cite{Clevert2015} function on all hidden layers. This activation was chosen as the method is aligned with the notion of natural gradients, which uses the geometry induced by Fisher Information to adjust the gradient direction \cite{Clevert2015}, pushing the activation to stay in a region that is not saturated.

The optimal number of layers (or residual blocks) generally depends on the complexity of the training data. Here, we use 10 residual blocks, which was found to provide the best results for the tests herein, comprising a total of $7913249$ parameters to be optimised.  For highly complex velocity structures, this number may need to increase.

The neural network learns the travel time between any source receiver pair and must contain an adequate sampling from across the 3D medium. The input features are organized into a vector of six components,
\begin{equation}
    \vec{x} = \left[X_s,Y_s,Z_s,X_r,Y_r, Z_r\right],
\end{equation}
where $X$, $Y$, $Z$ are the Cartesian coordinates of the source or receiver. The features are paired with the seismic velocity at the receiver location,
\begin{equation}
    y = V(X_r, Y_r, Z_r)
\end{equation}
A training dataset therefore consists of many $(\vec{x}, y)$ samples, taken from across the 3D volume (Figure \ref{fig:FlowDiagram}c). We discuss how these datasets are constructed in the following section.

The misfit between the predicted $\hat{V}$, as determined from equation \ref{eq:3}, and observed, $V$, velocities are then minimised using a mean-squared error loss function,
\begin{equation}
    L = \| V - \hat{V} \|^2.\label{eq:5}
\end{equation}
We use the Adam optimization algorithm for training with a learning rate of $5\times10^{-5} $\cite{Kingma2014} (Figure \ref{fig:FlowDiagram}d). The batch and dataset size are set to $752$ and $10^6$ respectively, with their variability discussed further in Section \ref{sec:btzsamp}. The dataset is sepearted into training and validation data, with the validation dataset at $10\%$ of the total size. An addition test dataset is created representing $10^4$ source-receiver pairs which are blind to the user prior to loss calculation. For all the simulations we use a single Nvidia Tesla V100 GPU, with models taking $66s$ per epoch given the parameters above. The effects of parameter values on computation cost can be found in Section \ref{sec:btzsamp}

Once trained the network can be applied to a series of user defined source-receiver pairs to determine the travel-time and predicted velocity, as shown in Figure \ref{fig:FlowDiagram}e.

\subsection{Building a training dataset} \label{methods:dataset}
Our approach builds a training dataset by randomly sampling source and receiver points from the continuous 3D medium and labeling each with the velocity at the receiver location. Since the velocity model is gridded, we use linear interpolation to map these values to a continuous domain. In this study, we explore three different methods for sampling the velocity model. In these formulations we investigate two sampling techniques from the classical theory of Bertrand paradox, sampling random points across the model space and sampling points at random distances.

\subsubsection{Random Distance}
The dataset is composed of a series of source locations selected randomly across the model space. Once a source point is selected, a receiver point in space is sampled at a random distance away from the source location along a random vector. This method inherently allows the source-receivers pairs to have a distance distribution that is uniform across the model space.

\subsubsection{Random Locations}
The dataset could also be composed of a series of randomly selected source and receiver locations. This allows for a more random distribution of points across the model space, but inherently has a non-uniform distribution of distances between source-receiver points, with a lower sampling at short distances, as expressed by the Bertrand Paradox.

\subsubsection{Weighted Sampling}
In complex velocity contrasting models the training procedure could quickly learn areas  of simple velocity variations, and we act to improve the training efficiency of this procedure by allowing dynamic resampling of the training source-point pairs for values of greatest misfit.
For the first epoch of training we minimise the misfit between the actual and predicted velocity estimates using a L2-norm, but also determine a importance weight parameter, $w$, for each training sample defined by: 
\begin{equation}
    w = \frac{\left\vert \hat{V} - V \right\vert}{V}.
\end{equation}
where $\hat{V}$ is the neural-network predicted velocity value and $V$ is the actual imposed velocity model. The weight value is high for the samples with the greatest relative misfit. For subsequent training epochs training samples are selected based on the weight value normalized by the maximum weight in the training dataset. To mitigate stagnation in the extremes of the weight distribution we project the weights between a user defined minimum and maximum, represented by $[\min,\max] = [0.1,0.9]$. This bound was chosen to mitigate the undersampling of regions with low misfit and oversampling of regions that could contain singularities.

\subsection{Model verification}
\begin{figure*}
\centering\includegraphics[width=0.75\textwidth]{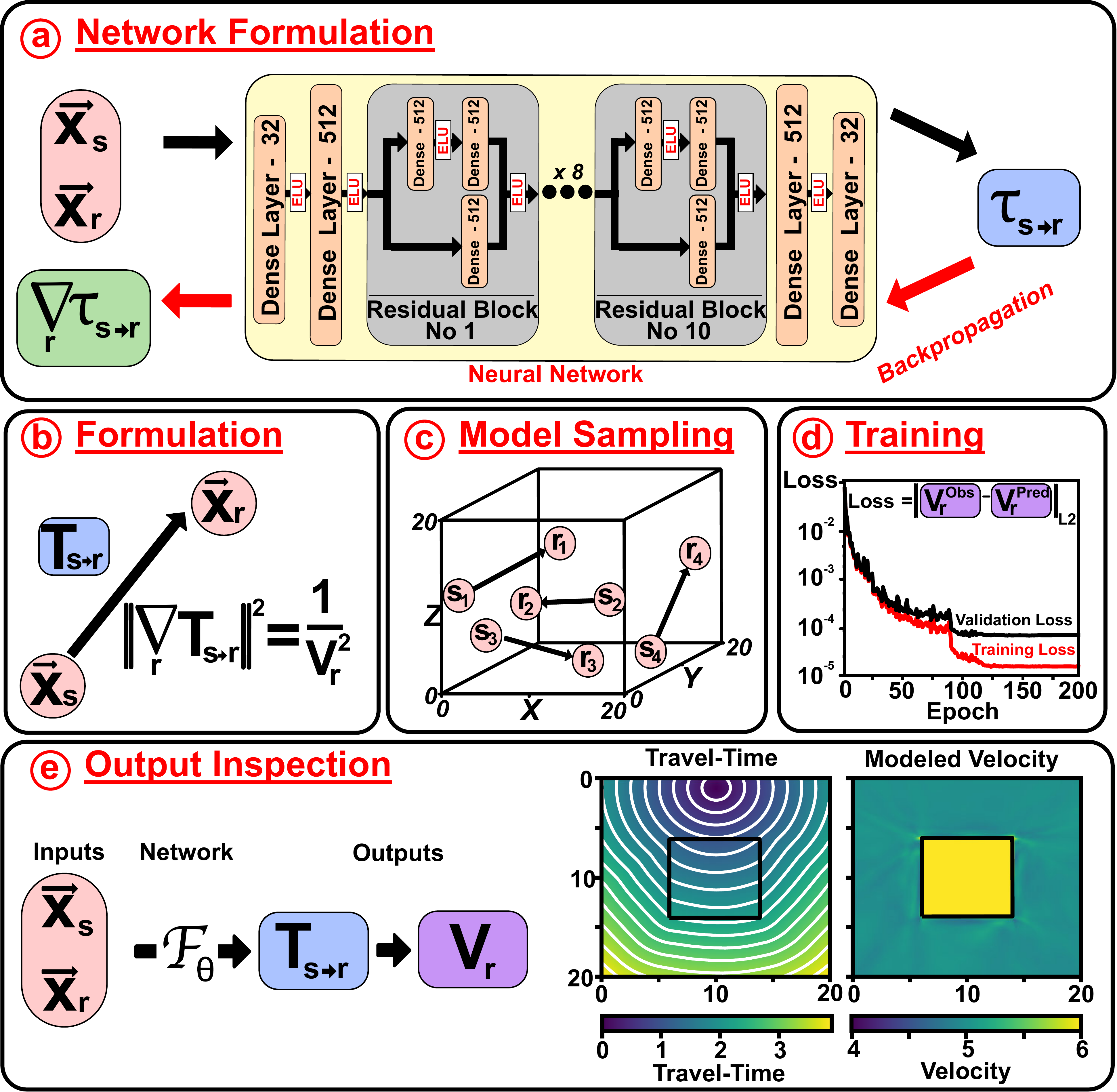}
\caption{Overview of processing workflow. (a) Neural network architecture composed of fully-connected layers and residual blocks. Each residual block is composed of 3 fully-connected layers with 512 neurons. ReLu activations are applied on all hidden layers. (b) Summary of Eikonal equation for $T_{s \rightarrow r}$ and $V_r$.  (c) Sampling of source-receiver pairs across the 3D volume to build the training dataset. (d) Network training through the minimization of loss function relating predicted and observed velocity values. (e) Inspection of neural network outputs by passing user defined source receiver pairs.}
\label{fig:FlowDiagram}
\end{figure*}

The accuracy of the solution to the Eikonal equation is given directly by the loss, which quantifies the degree to which the solution violates the PDE in a least squares sense. Thus, after training is finished, we can predict the travel time to all points desired within the 3D medium and use equation \ref{eq:1} to calculate the learned velocity model. Comparing the learned velocity model to the actual velocity model provides a visual and rigorous quantitative approach to understanding the accuracy of the solution (Figure \ref{fig:FlowDiagram}e).

\section{Baselines}\label{sec:FMM}
In this section we discuss some current approaches for solving the Eikonal formulation. While there have been many techniques developed for solving the Eikonal equation, a number of these can be broadly classified as either Fast Marching Methods (FMM) or Fast Sweeping Methods (FSM). The FMM is a grid based numerical scheme that uses a special finite difference operator to track the evolution of the minimum travel-time using an advancing interface scheme \cite{Sethian1996,Rawlinson2004}. In contrast, the FSM is an iterative method for solving the Eikonal equation via upwind differencing using a Gauss-Seidel iteration scheme \cite{Zhao2004}. This method groups the wavefronts by the sweeping directions in addition to meeting the requirement of increasing travel-times orthogonal to the wavefronts. Typically this sweeping procedure is implemented 6-times, for each of the positive and negative directions in the Cartesian coordinate system. As this method does not have to track an advancing interface, the computational cost is lower than the FMM, although this procedure can breakdown in the presence of strongly heterogenous velocity structures \cite{Capozzoli2014}. A more in-depth comparison between the FMM and FSM methods computational costs and solution misfit can be found in \cite{Capozzoli2014}.\\  

Throughout this study we utilize the python FMM as a baseline, since it is less sensitive to sharp velocity contrasts with only minor differences in the computational time to the Fast Sweeping Methods \cite{Capozzoli2014}. To compute the FMM travel-times we use the CPU toolkit scikit-fmm (\url{https://pythonhosted.org/scikit-fmm/}) to formulate the travel-time from a source location on a receiver geometry at $0.1km$ grid spacing in the $X,Y,Z$ dimensions. The root-mean squared (RMS) travel-time difference between the FMM and EikoNet approaches is calculated for each of the models. The FMM simulations are only used for comparison and are not used in the training of the neural network.

\section{Velocity Model Experiments}
Outlined are a series of experiments designed to demonstrate the versatility of our approach for learning travel time fields in complex 3D velocity models. We consider four different velocity models and examine the performance of the trained network. Each network is trained with a batch size of $752$, using a dataset with $10^6$ random distance source-points pairs with a weighting range of $[0.1,0.9]$. The effects of dynamic sampling and weighting are discussed further in Section \ref{sec:SamplingExp}.

\subsection{Homogeneous Velocity}
The first model we consider is a homogeneous 3D velocity structure with a value of $5km/s$ (Figure \ref{fig:SamplingType}a). For demonstration purposes, a source is placed at $\left[X,Y,Z\right] = \left[10,10,1\right]$, with both $X-Z$ and $X-Y$ slices of the travel time field shown. At each receiver point, we plot the learned velocity using equation \ref{eq:1}, and use this to determine the percent error. For comparison, we also show the FMM solution calculated on a grid of receivers with $0.1km$ grid spacing in the $X$,$Y$ and $Z$ dimensions, providing a travel-time root-mean-squared with the EikoNet travel-times. The training loss goes to zero after $110$ training epochs, showing that the network has learned the travel time field perfectly. When comparing the EikoNet solution with the analytical solution, the RMS travel-time error is zero. The FMM solution has a RMS travel-time error of $0.0113s$.

\subsection{Graded Velocity}
Next, we consider a velocity model that increases linearly with depth (Figure \ref{fig:SamplingType}b). The model increases from $3km/s$ at the surface, to $7km/s$ at $20km$ depth. Comparison with the conventional finite-difference scheme shows good agreement, with a mean difference between imposed and recovered velocity model of  $0.00209km/s$. The RMS travel-time with the FMM solution reaches a value of $0.0189s$.

\subsection{Block Model}
The third test conducted is a 3D model containing a central cube embedded within a homogeneous background (Figure \ref{fig:SamplingType}c). The cube has a constant velocity of $7km/s$, while the background velocity is $5km/s$. Here, the neural network approach has excellent misfit between the actual and learned velocity, with only minor disagreement close to sharp velocity gradients. The neural-network travel-time field is similar to that of the finite-difference scheme, with the effects of the sharp velocity contrast shown in the deflection of the travel-time fronts, a low travel-time RMS of $0.0311s$, and a low mean velocity difference of $0.094km/s$. This model demonstrats that the neural network approach is able to reconcile even difficult cases with sharp velocity changes of $20\%$ of the mean value.

\subsection{Checkerboard Velocity}
\begin{figure*}
\centering
\includegraphics[width=0.95\textwidth]{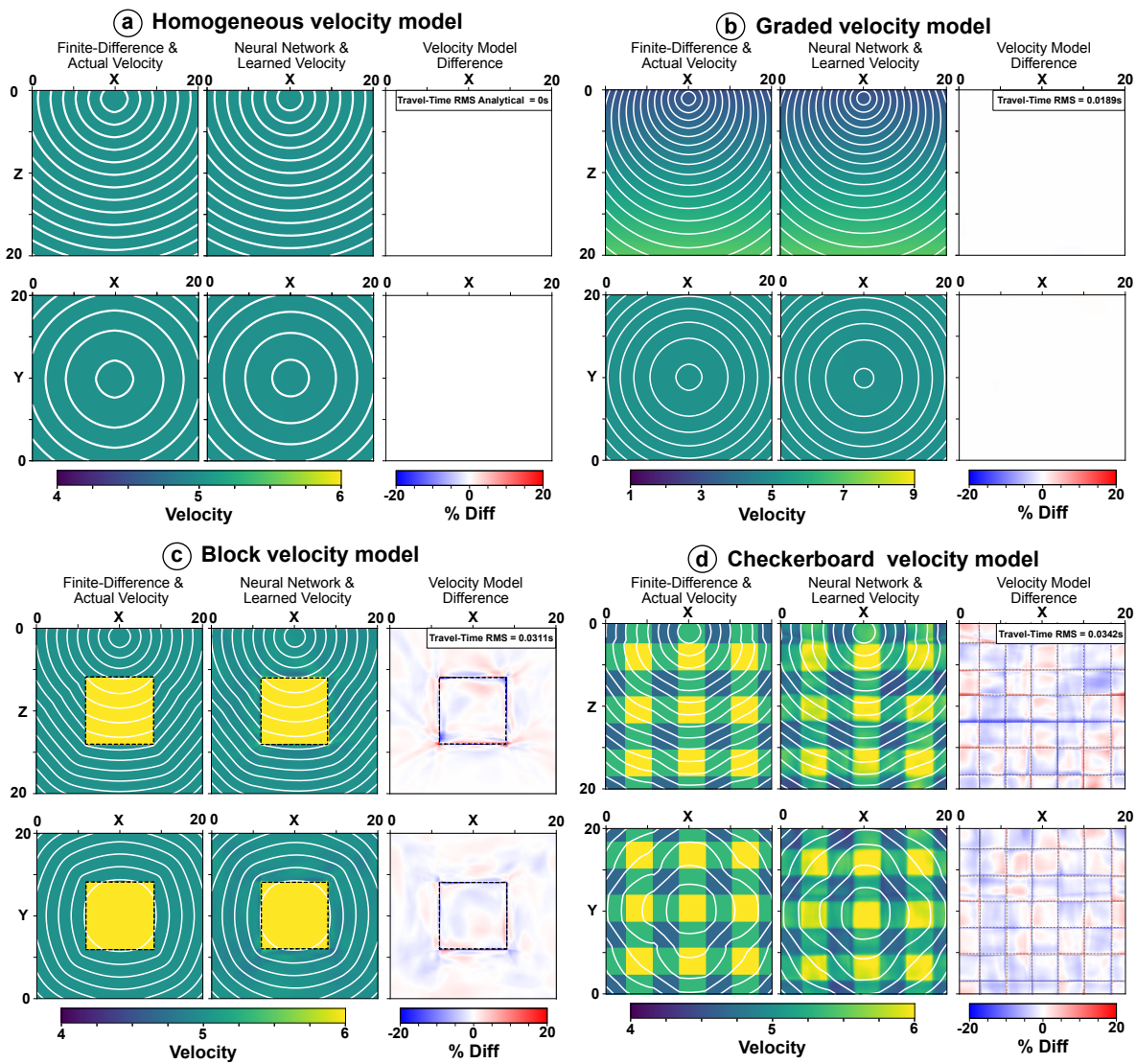}
\caption{Velocity model experiments with comparison to finite-difference and imposed velocity models. \textit{Left} panels represent the X-Z and X-Y slice from the imposed velocity model, overlayed by the finite-difference expected travel-time. \textit{Middle} panels represent the X-Z and X-Y slice from the neural network recovered velocity model and neural-network travel-time. \textit{Right} panels represent X-Z and X-Y slice velocity models differences between the imposed and recovered velocity model}
\label{fig:SamplingType}
\end{figure*}

The fourth test conducted contains a 3D checkerboard, which we use to demonstrate that the proposed algorithm is able to reconcile positive and negative velocity anomalies varying spatially within the domain. The velocity varies between $6km/s$ and $4km/s$, with a grid spacing of $6km$, meaning that the model is not symmetrical about the central point (Figure \ref{fig:SamplingType}d). We compare the solution with the finite-difference scheme and actual velocity model showing that the deep learning approach is able to reconcile the velocity model with a mean velocity difference of $0.19km/s$ and a travel-time RMS of $0.0342s$.

\subsection{Industry velocity Model - Marmousi2}
\begin{figure*}
\centering
\includegraphics[width=0.95\textwidth]{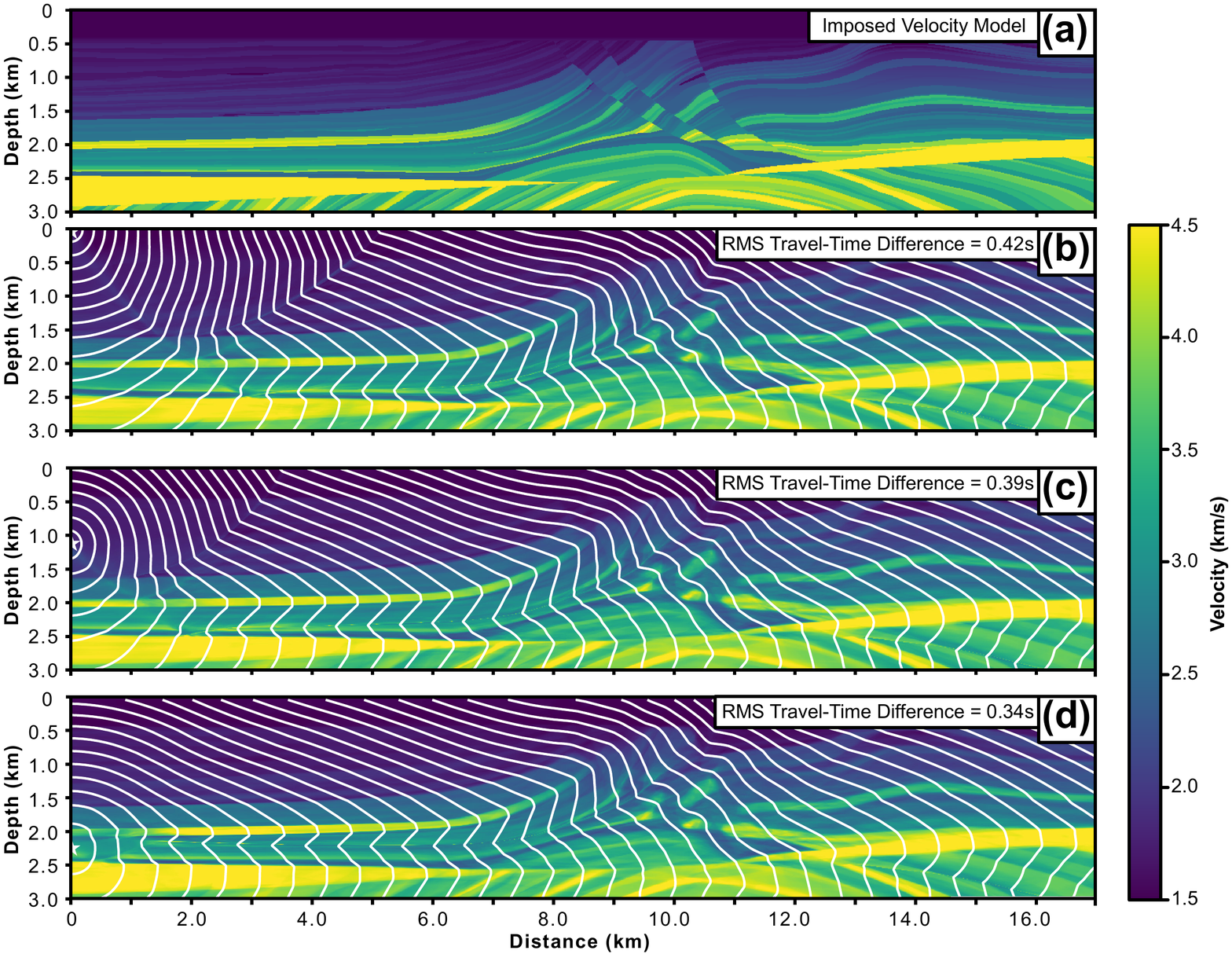}
\caption{Marmousi2 two-dimensional travel-time formulations using EikoNet and Finite-Difference Methods. (a) represents a colormap of the imposed 2D velocity model. (b) the recovered travel-time and velocity model fields from a source location at $[0,0]$ to a point grid at $0.001km$ spacing. Colormap represents the recovered velocity and white contours represent the travel-time at $0.1s$ spacing. (c) and (d) represent a similar plot to (b) but with the different source locations of $[0km,1.1km]$ and $[0km,2.25km]$ respectively.}
\label{fig:Marmousi2}
\end{figure*}

The final velocity model that we test is the Marmousi2 two-dimensional model. The Marmousi2, expanding on the original Marmousi model \cite{Versteeg1994}, is based on a geophysical profile through the North Quenguela Trough in the Cuanza Basin, Angola. This model represents a 17km by 3.5km cross-section of strongly heterogeneous velocity contrasts attributed to changes in subsurface geology, offsets from fault-structures and reservoir levels \cite{Martin2006}. We investigate the specific use of EikoNet on the S-wave velocity, as shown in Figure \ref{fig:Marmousi2}a. Due to the complexity in the velocity structure we expand our network to use 20 residual blocks, to enable EikoNet to encapsulate even the smallest resolution velocity contrasts. The network is trained across 400 epochs with sample points randomly selected for each batch to minimise the undersampling on sharp velocity contrasts. Figure \ref{fig:Marmousi2}a,\ref{fig:Marmousi2}b and \ref{fig:Marmousi2}c represent the EikoNet travel-time and velocity fields for a series of source locations, on a $0.001km$ grid spacing. For comparison FMM finite-difference travel-time simulations are created for each of the source locations on the same point grid spacing. Travel-time RMS is computed with values ranging between $0.34 - 0.42s$, and mean velocity model difference of $0.17km/s$. These simulations demonstrate that the EikoNet formulation is able to reconcile the sharp velocity contrasts of the complex subsurface geology, although some misfit still remains which could be improved in future work with more efficient sampling methods or improved network architectures.

\section{Sampling Experiments}\label{sec:SamplingExp}
Having demonstrated that deep neural networks can indeed learn to directly solve the Eikonal equation, we now examine the effect of the sampling scheme on the learned solution. Here, we use the block velocity model from the previous section and train separate models using each of the three sampling techniques described in Section \ref{methods:dataset}. We also separately test how the total number of training samples and batch size affect the performance.

\subsection{Sampling Schemes}
\begin{figure*}
\centering
\includegraphics[width=0.75\textwidth]{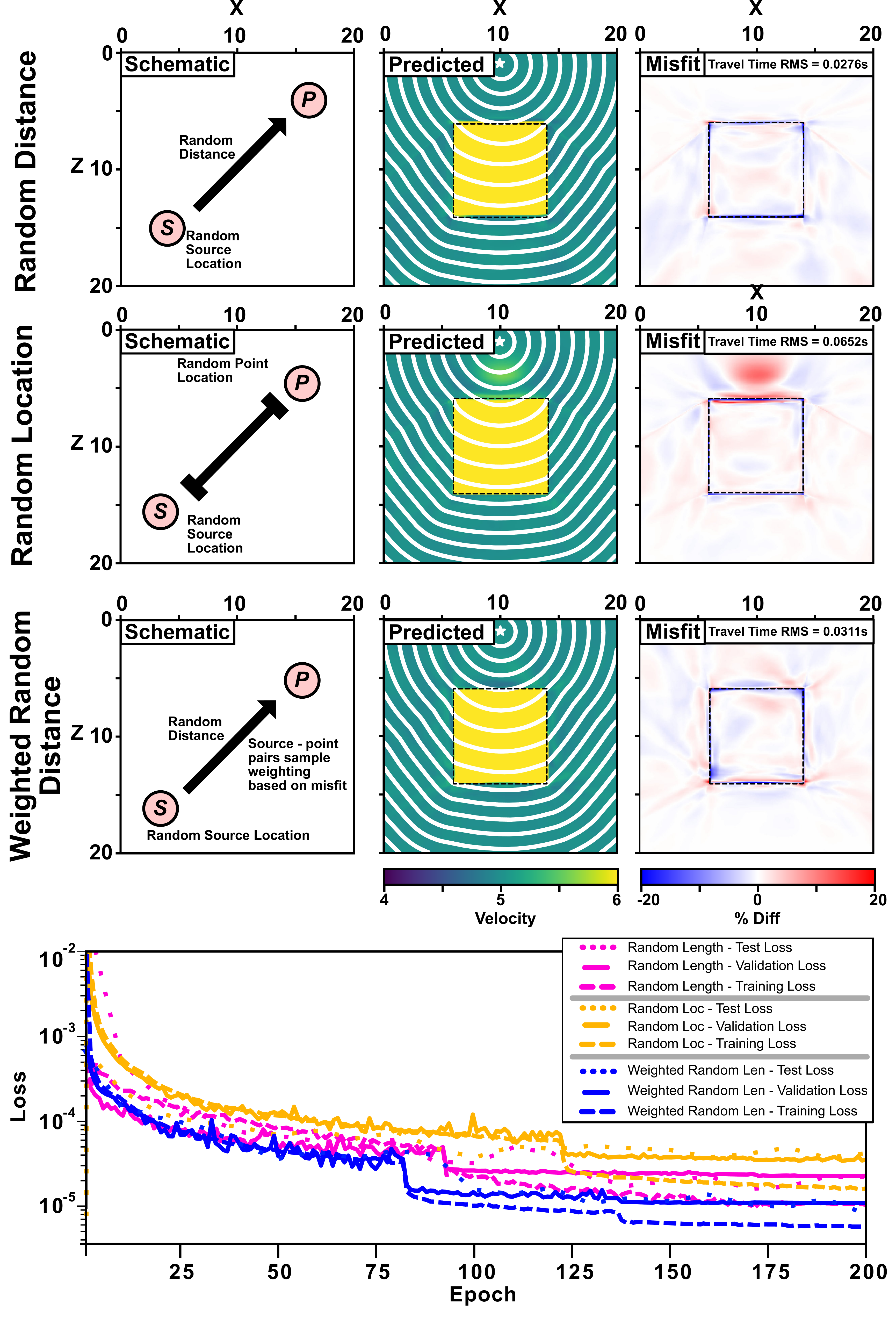}
\caption{Sampling schemes and their influence on the network performance. \textit{Left Column} represents a schematic of the different type of sampling, \textit{Middle Column} represents the learned velocity model for each simulation and \textit{Right Column} represents the misfit between predicted and imposed velocity model. \textit{Bottom Panel} represents the training, validation, and testing loss for each of the simulations.}
\label{fig:SamplingTypes}
\end{figure*}

Figure \ref{fig:SamplingTypes} shows the validation results for each of the three sampling methods. The weighted random sampling approach achieves the lowest loss of the three sampling methods, yet converges in a similar number of training epochs. However, the other two methods still perform well, as the differences in validation loss are relatively small. For the random location procedure there is greater misfit closer to the source location, expected due to the bias of the lengthscale to longer distances due to the Bertrand Paradox, and as such we use the random distance metric for sampling. Although the weighted sampler has little effect on the loss values and travel-time RMS value, we expect that the weighted random sampling will be of increasing importance in very complex 3D models and recommend it for selection of source-point pairs.

\subsection{Size of training dataset} \label{sec:btzsamp}
\begin{figure*}
\centering
\includegraphics[width=0.9\textwidth]{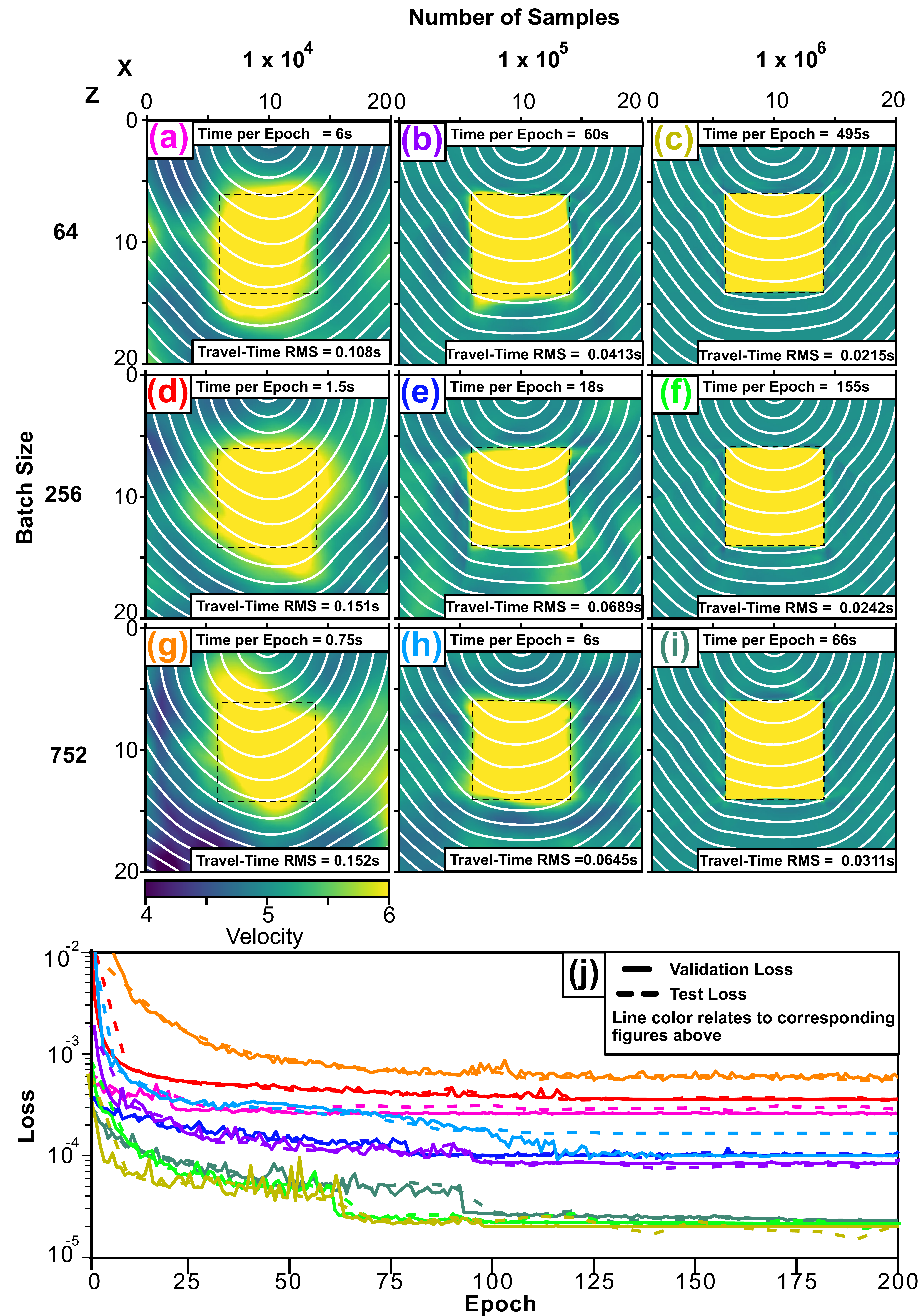}
\caption{Training loss effects with changing sample number and batch size. Panels (a) - (i) represent the different models runs, with number of samples increasing by columns left to right and batch-size increasing by rows from top to bottom. Panel (j) represents the validation loss, along with test loss for each of the separate model runs with color matching the panel labels.}
\label{fig:BtsizeSampleNum}
\end{figure*}

We investigate how the number of training samples and batch size affect the network performance (Fig. \ref{fig:BtsizeSampleNum}). We re-run the simulation with three dataset sizes ($10^4$, $10^5$ and $10^6$) and three batch sizes ($64$, $256$ and $752$), inspecting the recovery of the final solution and the validation loss for each simulation. Figure \ref{fig:BtsizeSampleNum}i shows the variation of the optimal recovered solution for the different batch and sample sizes, with columns representing the increasing number of samples and rows the increase batch size. Figure \ref{fig:BtsizeSampleNum}j shows the validation loss for each of the separate simulations with the line colour corresponding to the panel color.

From Figure \ref{fig:BtsizeSampleNum}, it is clear that the number of training samples has a profound influence on the network performance. The dataset with $10^6$ samples achieves a loss that is about an order of magnitude lower than the dataset with $10^4$ samples. In addition, through inspection of the recovered velocity model we can see that the low sample size is unable to reconcile the complex velocity structure of the sharp velocity contrast of the block model. Therefore, it is crucial that an adequate number of training samples number be used when constructing the dataset. Future work could investigate dynamic sampling of the velocity model space to reconcile regions of greatest misfit.

While the batch size is seen to influence the training results early on, the final best loss value is seen to be insensitive to this hyperparameter (Figure \ref{fig:BtsizeSampleNum}). This is important as larger batch sizes are much more computationally efficient.

\section{Future Applications}
In this section we discuss the application of EikoNet to a series of travel-time required problems, outlining the advantage of EikoNet over conventional Finite-difference methods and how these procedures would implemented.

\subsection{Earthquake Location}
In earthquake location theory a series of seismic instruments are used to record the arrival time of the incoming seismic wave. These instrument arrival times are then inverted using a velocity model to determine an earthquake location and location uncertainty. In recent years advances in seismological instrumentation have allowed for the incorporation of Distributed Accoustic Sensing (DAS), using optical fibres as a series of non-discretized station locations \cite{Williams2019}. Current travel-time finite-difference techniques are not tractable for handling the tens of thousands of virtual receivers that DAS arrays provide. This would require a large computational cost and disk storage space. In comparison our machine learning technique scales independently of the number of receivers, has a compact storage footprint, and can rapidly evaluate forward predictions from the network.

\subsection{Ray Multipathing}
\begin{figure*}
\centering
\includegraphics[width=1.0\textwidth]{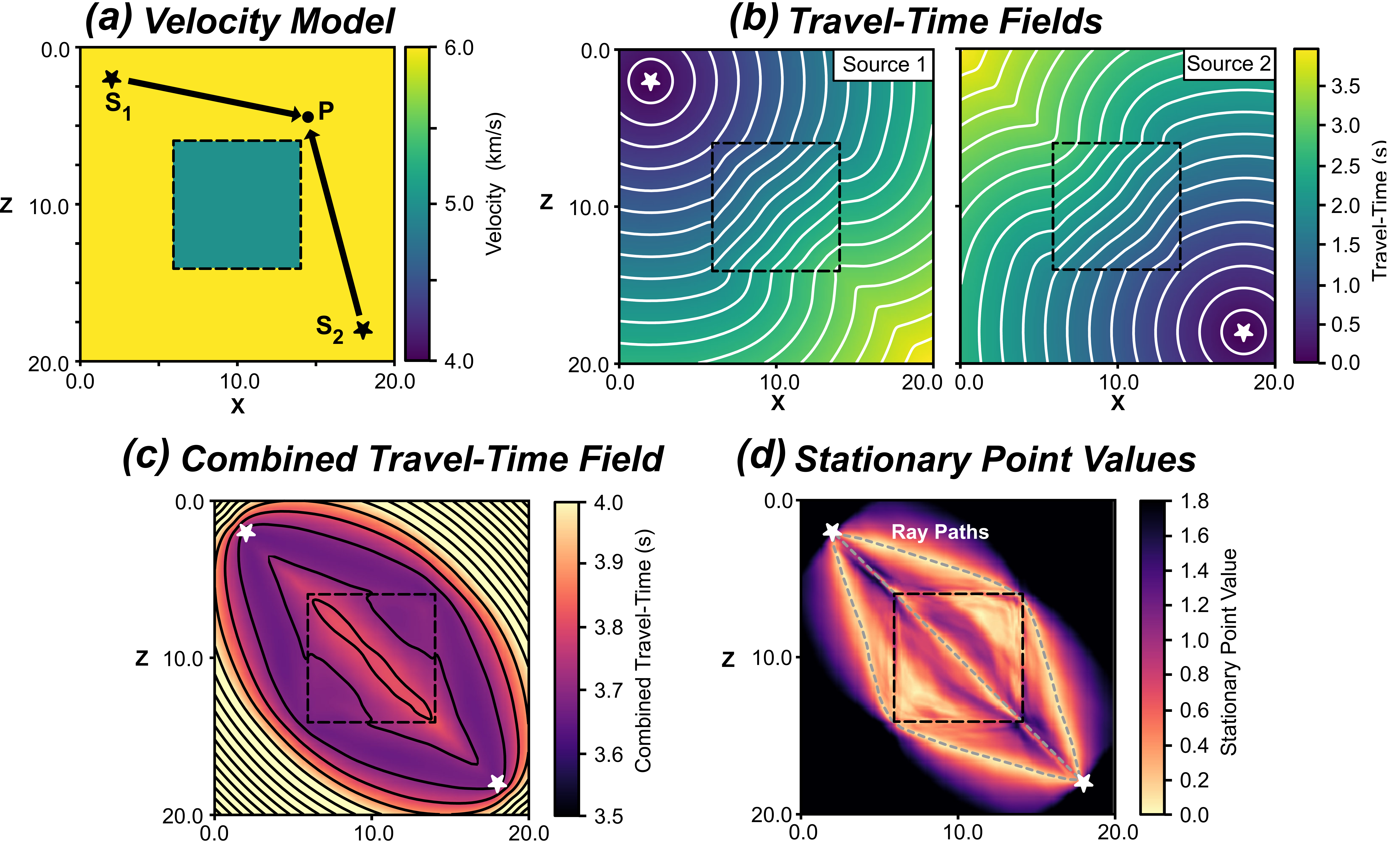}
\caption{Seismic ray multipathing procedure applied to an adapted block model velocity experiment with low velocity block and high velocity background. (a) Imposed velocity structure and schematic of the travel-times between source locations $S_1$ and $S_2$, to some point $P$. (b)  Travel-times from each of the receiver locations to a grid of receiver locations. (c) Combined travel-time field from the source locations to each of the receiver locations, $TT_P$. (d) Gradient of the combined travel-time field at each of the receiver locations, $\frac{\partial TT_P}{\partial P}$. Stationary points are represented where this value is zero, with these points representing possible secondary arrival ray paths.}
\label{fig:MultiPathings}
\end{figure*}

The Eikonal formulation represents the first arrival between source and receiver locations. However, in complex velocity models sharp gradients in the velocity structure can produce a multi-pathing effect with the energy partitioned between multiple ray paths. Rawlinson et al. \cite{Rawlinson2010} acted to mitigate this effect by employing FMM to track the evolution of the seismic wavefront for a narrow band of nodes, representing a interface of interest. Once the outgoing wave traverses one of these node locations an additional simulation is triggered and the combined wavefield used to determine a possible secondary pathway. 
The deep learning approach can be adapted to include additional secondary arrivals by determining the travel-time from two separate source locations to the same point in space. Figure \ref{fig:MultiPathings}a and \ref{fig:MultiPathings}b represent the travel-time field from the two source locations to a series of points within the block model velocity experiment, but now altered with a slow velocity centre and faster velocity background. By combining the travel-time for the source locations to each of the points we acquire the combined travel-time field as shown in Figure \ref{fig:MultiPathings}c. By taking the gradient of this combined-travel time across the network relative to the receiver locations (Figure \ref{fig:MultiPathings}d) we can determine if the points are stationary \cite{Rawlinson2010}, having a gradient value close to zero, and therefore representing a possible secondary phase arrival pathway. EikoNet is able to quickly formulate travel-time fields and take gradients relative to the receiver locations with each possible multipathing point calculated in $7s$ for $1\times10^6$ points on a single CPU, making this procedure beneficial for multipathing simulations.

\subsection{Tomographic Modeling}
\begin{figure*}
\centering
\includegraphics[width=0.8\textwidth]{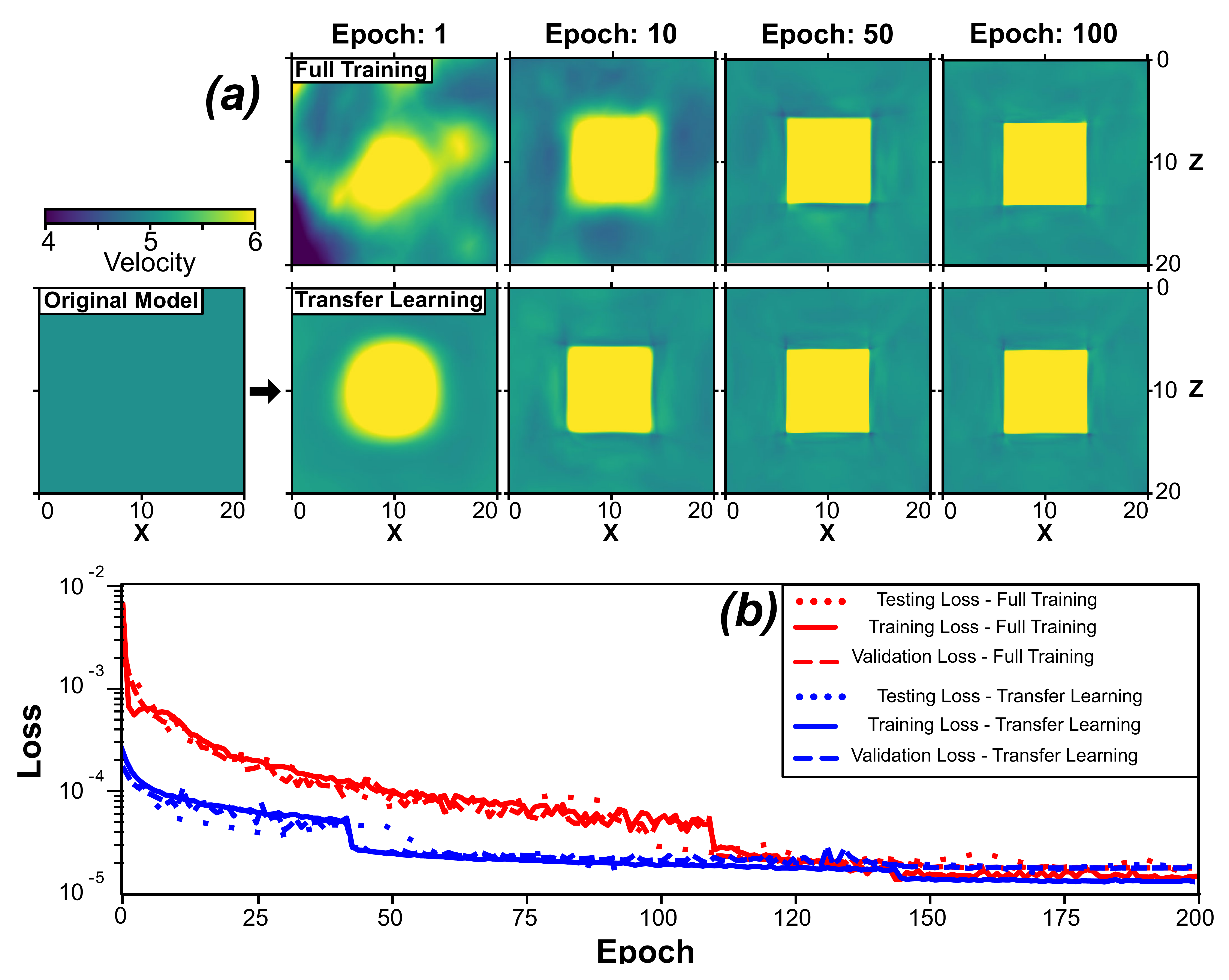}
\caption{Effects of Transfer Learning on the training loss for the Block model velocity experiment. (a) Training snapshots of a full-training procedure and transfer learning from Homogenous model. Rows represent the full-training and transfer learning approaches respectively. Columns represent the snapshot of the velocity model for a series of training epochs.}
\label{fig:TransferLearning}
\end{figure*}

For tomographic inversions which undergo many iterations successively, new travel-time fields must be computed from scratch for each iteration. Our approach allows for the neural network model from the previous iteration to be used as the starting point for the next training procedure, which could rapidly converge to the new velocity model if the perturbations are relatively small. This would effectively remove ray tracing as a computational burden from this part of the tomography, as nearly all of the compute time would be spent on the very first tomography iteration. We outline the use of a transfer learning technique on the Block model velocity experiment, comparing the training of a model from scratch relative to updating originally trained Homogeneous model. Figure \ref{fig:TransferLearning}a demonstrates the comparison of the returned models for a series of training snapshots. Figure \ref{fig:TransferLearning}b demonstrates that the transfer learning approach is $\sim 3-4\times$ faster than training a full model from scratch. 

The computation cost in the training procedure is in learning the complexities of the velocity model space. If the velocity model is an unknown but the user has some prior knowledge of possible arrival time differences, then this approach could be updated to do some form of tomographic inversion. This procedure instead would learn the velocity model to fit some known travel-time values. We perceive that this addition can be made in the loss function term itself, where an additional loss term can be used to update the velocity model to try and mitigate known observations. Prior finite-difference methods would have difficulty with this procedure as the travel-time field would have to be recalculated for each update to the velocity model.

\section{Discussion and Conclusion}
We have demonstrated that deep neural networks can solve the Eikonal equation to learn the travel-time field in heterogeneous 3D velocity structures. The method has been shown to produce solutions that are consistent with those of the Fast Marching Method.

Finite-difference approaches require computing the travel time field separately for every source location of interest, without any ability to pass along knowledge about the wavefield or velocity structure between simulations. The computation cost for the finite-difference approach therefore increases based on the number of source locations and receiver nodes, with the storage of these travel-time tables increasing drastically with grid size. The deep learning approach instead is able to learn from and generalize the knowledge acquired between multiple source locations, as much of the structure of the problem is highly similar. For example, if two sources are placed very close to each other, the travel-time field will be generally quite similar between them, and the information learned from solving the equation for the first source can be used to more rapidly learn the solution for the second source. This is not the case for fast-marching methods and only holds for small distances in fast-sweeping methods. Implicitly, this means that the neural network is learning the velocity model. The disk storage size required is equal to the size of the neural network ($\sim 90MB$ for 10 residual layers) as only the weights of the network have to be retained. For more complex velocity structures the neural network size may required a larger model, with the disk storage scaling by a linear function of the network size. However, for finite-difference approaches the storage size scales with the product of the number of source and nodal points, so for a complex velocity structure the model would require a fine-resolution grid spacing or dynamic meshing, making this method intractable with large large look-up table storage sizes. One of the distinguishing features of EikoNet is that the travel time solutions are valid for any two points within the 3D continuous domain. This means it is never necessary to store a travel time grid and interpolate it to achieve the desired result away from the grid nodes. Here, EikoNet automatically learns an optimized interpolation scheme during the training process, drawing on context from across the entire dataset. 

A second important aspect is that solutions to the Eikonal equation, as learned by EikoNet, are guaranteed to be differentiable back through the network with respect to the source or receiver locations. This has a variety of important practical applications, such as in locating earthquakes, as the inverse problem can be formulated as one of gradient descent, by analytically calculating the gradients of an objective function with respect to the source locations.

The Eikonal equation is not just used in seismology, but numerous other domains of wave physics such as optics \cite{Hoffnagle2011}, medical imagery \cite{Droske2001} and video-game rendering \cite{Huang2007}. It is expected that EikoNet would be just as suitable in these fields.

The computational cost of predicting the travel-time from a source to receiver location is equal to the time required to pass across the network ($4.047\times 10^{-4}s$ per source-point pair on a single 2.3GHz Intel Core i5 CPU). The low computation cost of a forward prediction means that the neural-network model can be used to significantly increase the computation speed of typically used ray-based procedures. Our approach is massively parallel and very well suited for GPUs (taking $0.424s$ for $1e6$ source-point pairs on a single Nvidia Tesla V100).

\newpage
\section*{Acknowledgment}
EikoNet is avaliable at github \hyperlink{https://github.com/Ulvetanna/EikoNet}{https://github.com/Ulvetanna/EikoNet}.\\

This project was partly supported by a grant from the USGS. K. Azizzadenesheli gratefully acknowledge the financial support of Raytheon and Amazon Web Services. We would like to thank Jack Muir for interesting discussions about finite-difference methods and limitations.


\bibliographystyle{plainnat}

\end{document}